# Extracting Geometry and Topology of Orange Pericarps for the Design of Bioinspired Energy Absorbing Materials

*Chelsea Fox[1], Kyle Chen[1], Micaela Antonini[2], Tommaso Magrini[1,3]\* and Chiara Daraio[1]\**

[1] Division of Engineering and Applied Science, California Institute of Technology, Pasadena, CA 91125, USA

[2] Department of Biotechnology and Life Sciences, University of Insubria, Varese, Italy

[3] Department of Mechanical Engineering, Eindhoven University of Technology, 5600MB Eindhoven, The Netherlands

\* Email: t.magrini@tue.nl, daraio@caltech.edu

## Keywords

Additive manufacturing, Irregular materials, Bioinspired materials, Energy absorption

## Abstract

As a result of evolution, many biological materials have developed irregular structures that lead to outstanding mechanical properties, like high stiffness-to-weight ratios and good energy absorption. To reproduce these properties in synthetic materials, biomimicry typically replicates the irregular natural structure, often leading to fabrication challenges. Here, we present a bioinspired material design method that instead reduces the irregular natural structure to a finite set of building blocks, also known as tiles, and rules to connect them, and then uses these elements as instructions to generate synthetic materials with mechanical properties similar to the biological materials. We demonstrate the method using the pericarp of the orange, a member of the citrus family known for its protective, energy-absorbing capabilities. We generate polymer samples and characterize them under quasi-static and dynamic compression and observe spatially-varying stiffness and good energy absorption, as seen in the biological materials. By quantifying which tiles and connectivity rules locally deform in response to loading, we determine how to spatially control the stiffness and energy absorption.

**Introduction**

Nature provides many examples of materials with desirable mechanical properties, such as high strength[1–5], high toughness[4,6–8], and high impact resistance[2,9–12]. Some of these materials are periodic, like nacre[1,4], conch shells[13], and beetle wings[5], while others have irregular structures, like trabecular bone[2,9,10] and citrus pericarp [14,15]. However, periodic bioinspired materials are more widely studied than irregular biological materials, as they are more easily fabricated via additive manufacturing[16–21] and studied computationally[18,19,21–24]. Conversely, generating irregular materials often requires complex biomimicking processes such as micro-computed tomography coupled with 3D printing[25], or investment casting[26,27]. Other approaches included the use of stochastic processes, such as a virtual growth algorithm[28,29], Voronoi tessellations[30–32] and foaming[33–37] for irregular materials generation, but these methods are limited in their ability to imitate the biological structure. Indeed, irregular biological materials are often defined by highly complex and geometrically irregular concave and convex internal structures[3,11,14,38,39], as well as spatial density variation, optimized to respond to specific loading conditions[2,15,25,39,40].

Here, we propose a bioinspired material design method that instead reduces the irregular biological structure to a finite set of tiles and rules for how to connect them, and then uses these elements as instructions to generate synthetic materials. It has been shown that topology and geometry control the mechanical properties of materials, such as elastic modulus, shear modulus, strength and Poisson's ratio[41–52], and the finite set of tiles and connectivity rules are analogous to topology and geometry, respectively. Each tile has an associated topological coordination number, R, which is defined as the number of branches from the central node[53], while the connectivity rules define the geometry by determining how tiles fit together to create the irregular structure. Tiling and tessellation approaches that generate irregular geometries have already been studied to achieve mechanical properties such as stiffness and strength[28,54–58], but these approaches often focus on homogenization or have very limited size due to computational costs[59], whereas our approach seeks to quantify the structure and its mechanical properties at a local tile level that can be spatially controlled and scaled.

To demonstrate the bioinspired material design method, we focus on the pericarp of the orange, a member of the citrus family[38]. Citrus fruits are known for their thick pericarps, which range from 5-7 mm for oranges and lemons to 15-20 mm for the citron (the thickest pericarp)[60]. Regardless of the type of fruit, these thick pericarps consist of an irregular, density-graded foam-like structure, which has evolved for energy absorption and impact resistance, key to protecting the pulp when the ripe fruits fall from the tree[15,60]. The dense outer layer of the pericarp, known as the flavedo, acts as a protective layer, while the less dense internal region, known as the albedo, provides energy absorption due to the presence of large, compressible intercellular spaces[15]. Furthermore, vascular bundles throughout the structure act as reinforcing elements, providing additional strength and stiffness[15]. We determine the tiles and connectivity rules of the orange pericarp and then use these as instructions to generate synthetic samples with the same structure as the fruit through a computer-aided virtual growth algorithm[28] (VGA), which we then additively manufacture. Under quasi-static and dynamic compression, we observe spatially-varying stiffness and energy absorption similar to that of the biological material, indicating that the tiles and connectivity rules are sufficient structural descriptors to imitate the mechanical properties. We then quantify which tiles and connectivity rules produce a particular property by examining the local deformation to understand how to spatially control the mechanical performance.

**Methods, Results, and Discussion**

*Bioinspired Material Design*

We begin with a two-dimensional cross-sectional image of an orange pericarp, acquired transversely from the external surface to approximately 5 mm into the fruit, where the pericarp transitions into the pulp (Figure 2a, Figure S1, Supporting Information). We use the image processing software, FIJI[61], to skeletonize[62] the structure into a simplified line form that maintains the topology and the complex geometry of the original pericarp image (Figure 1a). This irregular image skeleton is then broken down using a uniform square grid into a collection of tiles (Figure 1b,c). The tile size is determined by taking the largest possible size while ensuring that each tile contains no more than one node, defined here as an intersection point between branches (Figure 1b). Although each tile contains a unique portion of the original image of the orange pericarp, all tiles can be translated into a finite set of tiles with a reduced geometry. We perform the translation process between the original tiles and the reduced geometries by analyzing the perimeter of each tile and counting the number of times that a tile branch intersects the perimeter and assign a binary code value to the left (1), top (10), right (100), and bottom (1000), or assign zero otherwise (Figure 1d). By summing the perimeter values for each unique biological tile, we can determine its coordination number[63], as well as its orientation (Figure 1f, Supporting

Information Discussion 1). From the perimeter and coordination analysis, we then generate the finite set of reduced geometry tiles (Figure 1d). Next, we determine the connectivity rules governing how the tiles are assembled by examining the frequency at which two tiles are adjacent in the biological structure (Figure 1e,g, Supporting Information Discussion 1). The frequency of the reduced geometry tiles and the rules that determine their connectivity, two parameters that can be extracted from any starting structure (Figure 1h), are then supplied to a virtual growth algorithm[28] to assemble the irregular bioinspired structure (Figure 1i).

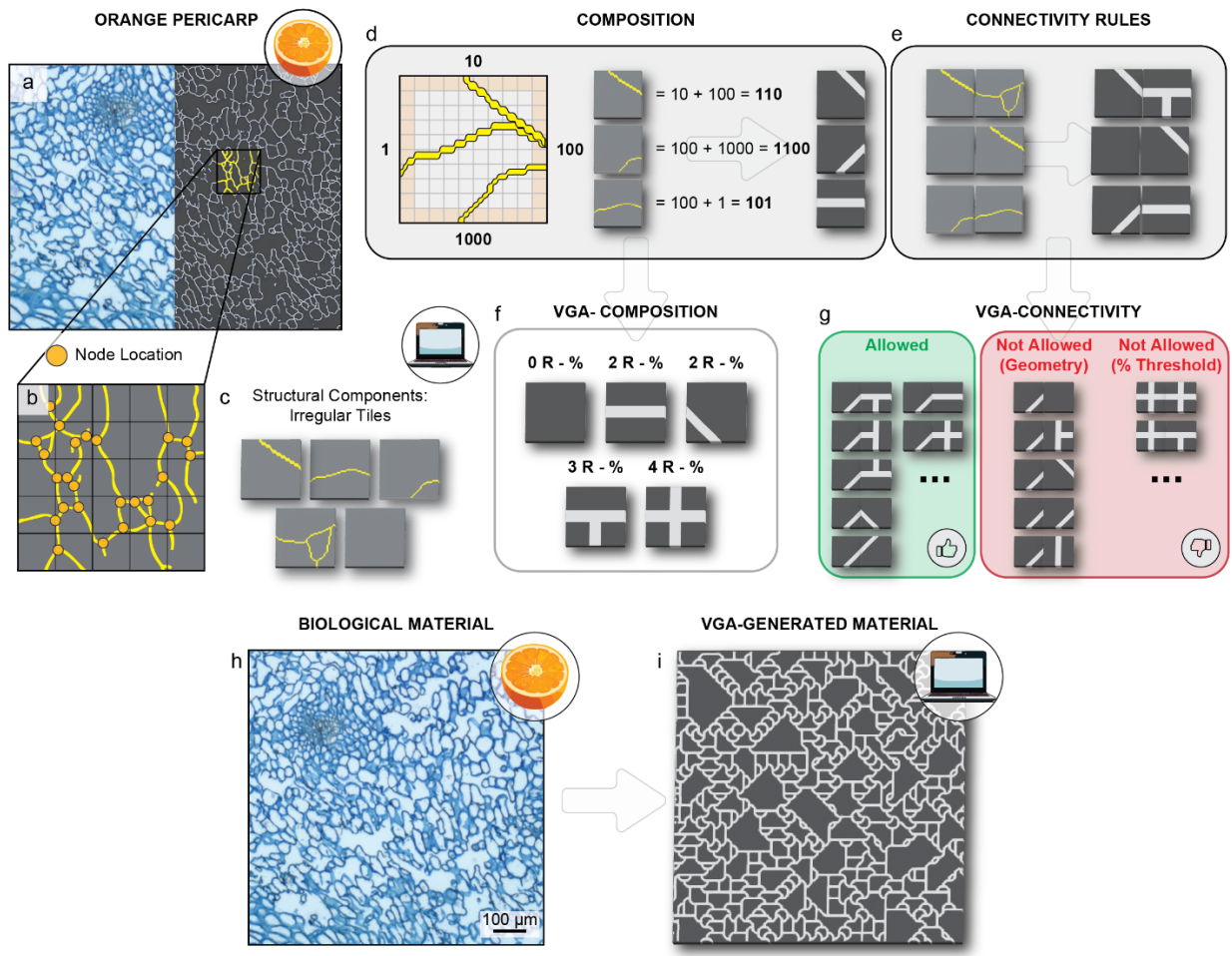

**Figure 1.** Bioinspired material design method. (a) Orange pericarp image and resulting skeletonized structure. (b) Skeletonized structure with node locations. (c) Irregular tiles from skeleton. (d) Translation of biological tiles into reduced geometry tiles using perimeter identification numbers and summation. (e) Connectivity rules from skeleton image. (f) VGA tiles and their coordination number. (g) VGA-allowed and -not-allowed connectivity rules. (h) Example image of orange pericarp. (i) VGA-generated sample.

*Material Generation and Fabrication:*

To generate samples, we divide the orange pericarp into external and internal regions, which correspond to the flavedo and the albedo[15,38], respectively (Figure 2a,b,c). The external region is taken from 0 mm to 0.5±0.16 mm into the pericarp and the internal region is taken from approximately 0.5±0.16 mm up to 5 mm into the pericarp, where the endocarp and pulp begin. Using the gridded approach on eight different equivalent pericarp images (Figure S1, Supporting Information), we determine that the external region is composed of 20.6 ± 1.3% of coordination number zero (0-R) tiles, 34.8 ± 1.2% of coordination number two (2-R) tiles, 34.7 ± 1.4% of coordination number three (3-R) tiles, and 10.0 ± 1.0% of coordination number four (4-R) tiles (Figure 2d, blue). By analyzing the connections between the tiles in the external region, we determine that all connectivity rules should be included except between 0-R tiles and 0-R tiles, and between 4-R tiles and 4-R tiles, which appear below the connectivity threshold (Figure 2e, blue, Supporting Information Discussion 1). The internal region is composed of 26.6 ± 1.6% of 0-R tiles, 36.9 ± 1.3% of 2-R tiles, 28.2 ± 1.4% of 3-R tiles, and 8.3 ± 0.9% of 4-R tiles (Figure 2d, orange). It is noted that larger intercellular spaces present in the internal regions of the orange pericarp lead to a significantly higher concentration of 0-R tiles (Figure 2d, orange). Furthermore, in the internal region, all connectivity rules are included except those between 2-R tiles and 4-R tiles, 3-R tiles and 3-R tiles, and 4-R tiles and 4-R tiles (which also occurs below the connectivity threshold) (Figure 2e, orange, Supporting Information Discussion 1).

With the tile frequencies and connectivity rules extracted from the two different regions of the orange pericarp, a computer-aided virtual growth algorithm[28] uses the set of reduced geometry tiles to generate 50x50 tile samples for the external (VGA-ext) and the internal (VGA-int) regions (Figure 2f,g). These samples are then combined together to form the bioinspired equivalent of the orange pericarp (VGA-full): a continuous structure with interface-free, spatially-varying density and structural features, defined as the area enclosed by cell walls (Figure 2h). We 3D print the VGA-generated geometries into a two-phase composite material to create a structure suitable for mechanical testing, image analysis and strain mapping. We use a polyjet printer (Stratasys Objet500 Connex3), with a stiff viscoelastic resin (Stratasys VeroWhite Polyjet Resin) for the reinforcing structure, and a soft elastomeric resin (Stratasys TangoBlack Polyjet Resin) for the matrix, both of whose mechanical properties fall within those reported in literature[63–65].

Although we exactly imitate the topology of the original sample by matching tile percentages, we must also quantify how well the VGA-generated samples' geometry compares with the original orange pericarp. For this, we use two different metrics: density, as a measure of structural feature size, and concavity, as a measure of structural feature shape. We observe that the VGA-generated samples maintain the same density difference between the external and internal regions as the orange pericarp samples, with 3-5% lower density in the internal region as a result of the larger structural features (Figure 2i). To compare the concavity of the original samples with the VGA-generated samples, we examine each structural feature, using the bridge length to Euclidean distance ratio, where the bridge length is the length of a structural feature's edge between 3-R and 3-R or 4-R nodes, and the Euclidean distance is the linear distance between the 3-R and 3-R or 4-

R nodes (Figure S2, Supporting Information). A ratio value of 1 indicates no concavity, whereas a higher ratio indicates a greater degree of concavity. When compared with the orange pericarp's structural features, the VGA-generated samples are very similar, with average concavity values within 2% for both regions, and with the internal region features having significantly higher concavity than the external region features (Figure 2j).

We also observe that the orange pericarp and the VGA-generated samples are isotropic, which can be shown by examining the orientation of the structural features averaged over several samples to determine how the features are distributed. Using the orientation of an elliptic fit in MATLAB (MathWorks, USA), we observe a uniform distribution of structural feature orientations in the internal region, while the distribution is more bimodal (0° and 180°) in the external region (Figure S3, Supporting Information). This bimodal distribution is due to the extensive presence of small, approximately circular cells, but it is an artefact, as circular cells are inherently isotropic. Although the angles of orientation show a uniform distribution, they are not an independently sufficient metric to confirm an isotropic material distribution, because they do not consider the effect of feature size. Therefore, we also verify that the orientation angles do not correspond to a certain structural feature size to ensure isotropy. There is no correlation between size and angle in the orange pericarp internal and external regions, or in the VGA-int samples, although there is a slight correlation for angles of 0°, 45°, 90°, and 135° in the VGA-ext samples, due to the four-sided nature of the virtual growth algorithm coupled with higher coordination number resulting in lower polydispersity (Figure S3, Supporting Information). Finally, it should be noted that the internal region VGA samples *individually* are not isotropic because their mechanical response is dominated by a few of the largest features, due to size limitations for testing (Figure S4, Supporting Information).

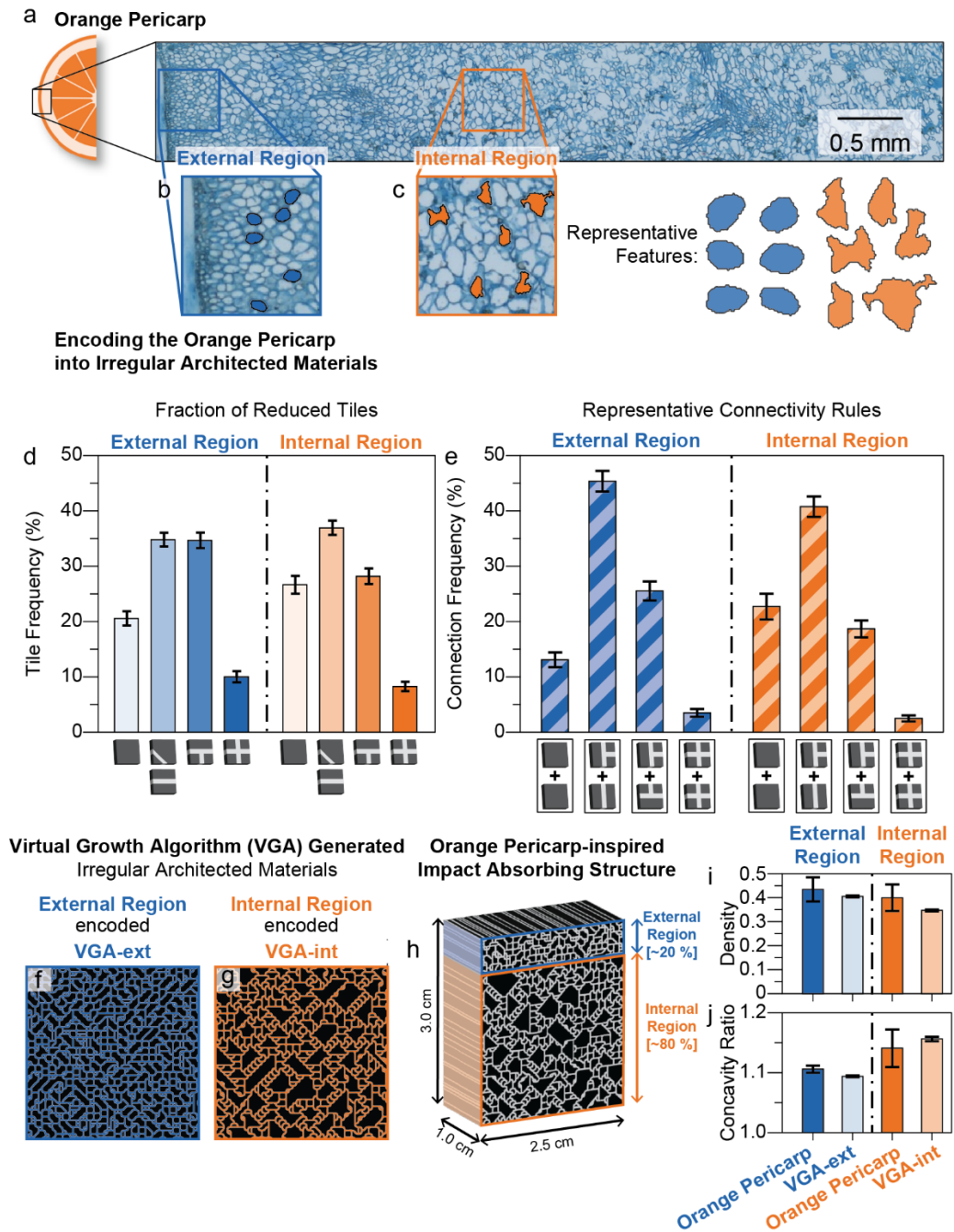

**Figure 2.** Orange pericarp characterization and VGA sample generation. (a) Cross-sectional image of orange pericarp. (b) Representative image of external region. (c) Representative image of internal region. (d) Tile percentages for external and internal regions. (e) Selected connectivity rule

percentages for external and internal regions. (f) VGA-generated sample of external region. (g) VGA-generated sample of internal region. (h) 3D-printed composite polymer sample with 80% + 20% external and internal regions, respectively. (i) External and internal region density for orange pericarp and VGA samples. (j) External and internal region concavity ratio for orange pericarp and VGA samples.

*Mechanical Characterization: Quasi-static Testing*

After establishing the equivalence of the topology and geometry of the VGA-generated samples with the original orange pericarp samples, we conduct quasi-static compression tests on the additively manufactured polymer composite samples (Supporting Information Discussion 2). Like the original orange pericarp, which features a stiff, protective flavedo[15,38], VGA-ext samples prove to be 116% stiffer than the VGA-int samples (Figure 3a,b, respectively) and even when normalized for their difference in density, VGA-int samples prove more compliant, like the energy-absorbing albedo of the orange pericarp[14,15,60]. As expected from a rule of mixtures argument, the orange pericarp inspired samples, VGA-full, featuring a 20% external and 80% internal composition, have a stiffness that is intermediate between the VGA-ext and VGA-int samples (Figure 3c). The constitutive stress-strain plots for VGA-full samples with a different composition, featuring 10% external and 90% internal, as well as 50% external and 50% internal, display the same rule of mixtures trend (Figure S5, Supporting Information). We can also quantify which tiles and connectivity rules are primarily responsible for the stiffness variations by examining the local sample deformation. We quantify the strain field experienced by each sample up to 10% total strain, using 2D digital image correlation (2D DIC) to identify which features are the stiffest (and undergo the least deformation) and which features are the least stiff (and undergo the most deformation) (Figure 3d,e,f). 2D DIC shows that the strain field in VGA-ext samples is more uniformly distributed, with no region exceeding 15% local strain at a global strain of 10% (Figure 3d, III) and that the samples are more homogenously composed of many smaller stretching-dominated[29] structural features (Figure 3d), like the protective flavedo and reinforcing vascular bundles of the orange pericarp[15]. These features are formed by higher percentages of high coordination tiles and by limiting large consecutive 0-R and 2-R tile connections, which results in local structural feature coordination numbers from ~2 up to 3.5, and lower polydispersity (Figure 3d, III, insets). In contrast, VGA-int samples display a significantly more localized strain field, with certain structural features reaching up to 30% local strain, although these local regions of high strain are uniformly distributed across the sample (Figure 3e, III, insets). We can observe that the VGA-int samples are composed of larger structural features, like the highly compressible, large intercellular spaces of the orange pericarp[15], formed by higher percentages of low coordination tiles and large consecutive 0-R tile connections as well as by consecutively aligned tiles (such as 2-R to 2-R, or 3-R to 3-R with the same orientation) or by consecutively repeating two-tile combinations (such as repetitions of the same 2-R to 3-R pair), which prevents the diversion of the feature edges (Figure 3e, III, insets). This results in local coordination numbers as low as ~1.5, allowing for less stiff bending and buckling mechanisms[50], as well as higher polydispersity (Figure 3e, III, insets). Furthermore, the largest, most deformed structural features have many concave edges, formed by the connection of diagonal 2-R tiles with 3-R tiles, which act as less-constrained joints[47,48] that can rotate as the feature deforms (Figure 3e, III, insets). Finally, the VGA-full sample shows the same trends, with the internal region displaying highly localized strain values up to 30% (Figure 3f, III, insets), while the external region never exceeds 15% local strain. It is also important to note that the two regions are smoothly connected and do not experience any additional strain localization in the region where the external region transitions to the internal region.

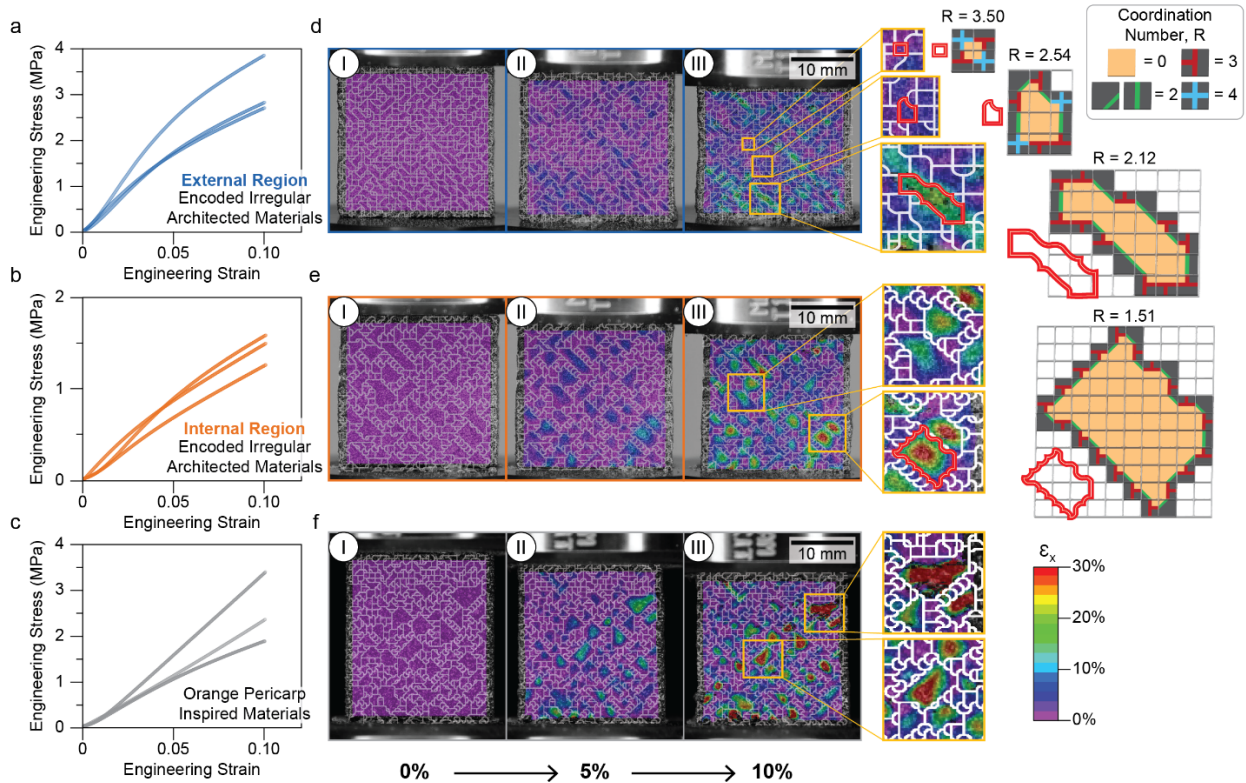

**Figure 3.** Quasi-static compression tests. (a) VGA-ext engineering stress-strain plot for three different samples. (b) VGA-int engineering stress-strain plot for three different samples. (c) VGA-full engineering stress-strain plot for three different samples. (d) Example 2D DIC strain maps for 0%, 5%, and 10% global strain for VGA-ext sample, yellow insets show coordination color coded example structural features. (e) Example 2D DIC strain maps for 0%, 5%, and 10% global strain for VGA-int sample, yellow insets show concave structural features at high strain and coordination color coded example structural feature. (f) Example 2D DIC strain maps for 0%, 5%, and 10% global strain for VGA-full sample, yellow insets show concave structural features at high strain.

*Mechanical Characterization: Dynamic Testing*

To characterize the energy absorption capabilities of the VGA-generated samples, drop tower tests at a strain rate of ~100 s$^{-1}$ are conducted on polymer composite samples that are equivalent to those tested under quasi-static conditions. To quantify the energy absorption capabilities of the VGA-generated materials, we measure the time of contact between the striker and each sample, and the coefficient of restitution, defined as the ratio of the average velocity of the striker in the 2 ms before and after impact with the sample. The velocities are measured using the image processing software, FIJI[61], by reslicing a vertical line drawn through the center of each image of the event (Figure 4a, vertical dashed line). Evaluating the evolution of the pixel values over the duration of the experiment allows us to track the position of the striker over time, using the incoming and outgoing angles defined by the position versus time resliced image of the striker and sample midline (Figure 4b). All VGA-generated samples are tested along with a periodic honeycomb sample with the same volume fraction of reinforcing phase as the 20% + 80% VGA-full sample, for comparison. After normalizing for density, we observe that the VGA-ext samples have the highest coefficient of restitution (0.45±0.03), indicating the least amount of energy dissipated (Figure 4c), along with the shortest time in contact (Figure 4d). Despite having the same volume fraction of reinforcing- and matrix phases, the VGA-full samples have a 7.5% lower coefficient of restitution (0.37±0.02) than their periodic equivalents (0.40±0.01) as well as a 16% longer time in contact. We also test the same 10% and 90% as well as 50% and 50% combination samples of external and internal regions to compare the coefficients of restitution and observe a linear trend of decreasing internal region thickness with an increased coefficient of restitution (Figure S5, Supporting Information). To explain these differences in energy absorption, we observe a positive correlation between structural feature size, overall sample concavity and amount of deformation, noting that higher concavity leads to less constrained strut bending and buckling in the largest structural features (Figure 4e,f), like the large, compressible intercellular spaces of the orange pericarp, known for their energy dissipation[15]. As previously discussed, the high percentages of low coordination tiles and consecutively repeating aligned tiles or two-tile combinations are responsible for the large size, while the connection of diagonal 2-R tiles with 3-R tiles are responsible for the high concavity (Figure 4g). Indeed, the largest example structural features have local coordination numbers as low as 1.28 (Figure 4f, I) and concavity ratios as high as 1.17 (Figure 4f, IV). We can also observe by plotting the bending angle of an example large structural feature over the duration of an experiment that the increase in angle follows a non-linear trend, as supporting adjacent features do not deform homogenously in time due to the irregularity of the geometry (Figure 4h). This results in isotropic global deformation, unlike periodic materials, like the honeycomb, which deforms anisotropically[32]. To quantify the strain during deformation, we can also refer back to the quasi-static 2D DIC maps (Figure 3d,e,f), which are valid also at higher strain rates since the striker velocity is between 7-8 m/s and the elastic wave speed in the material is approximately 575 m/s, indicating that drop tower loading occurs slowly enough to reach a state of stress equilibrium (Figure S6, Supporting Information).

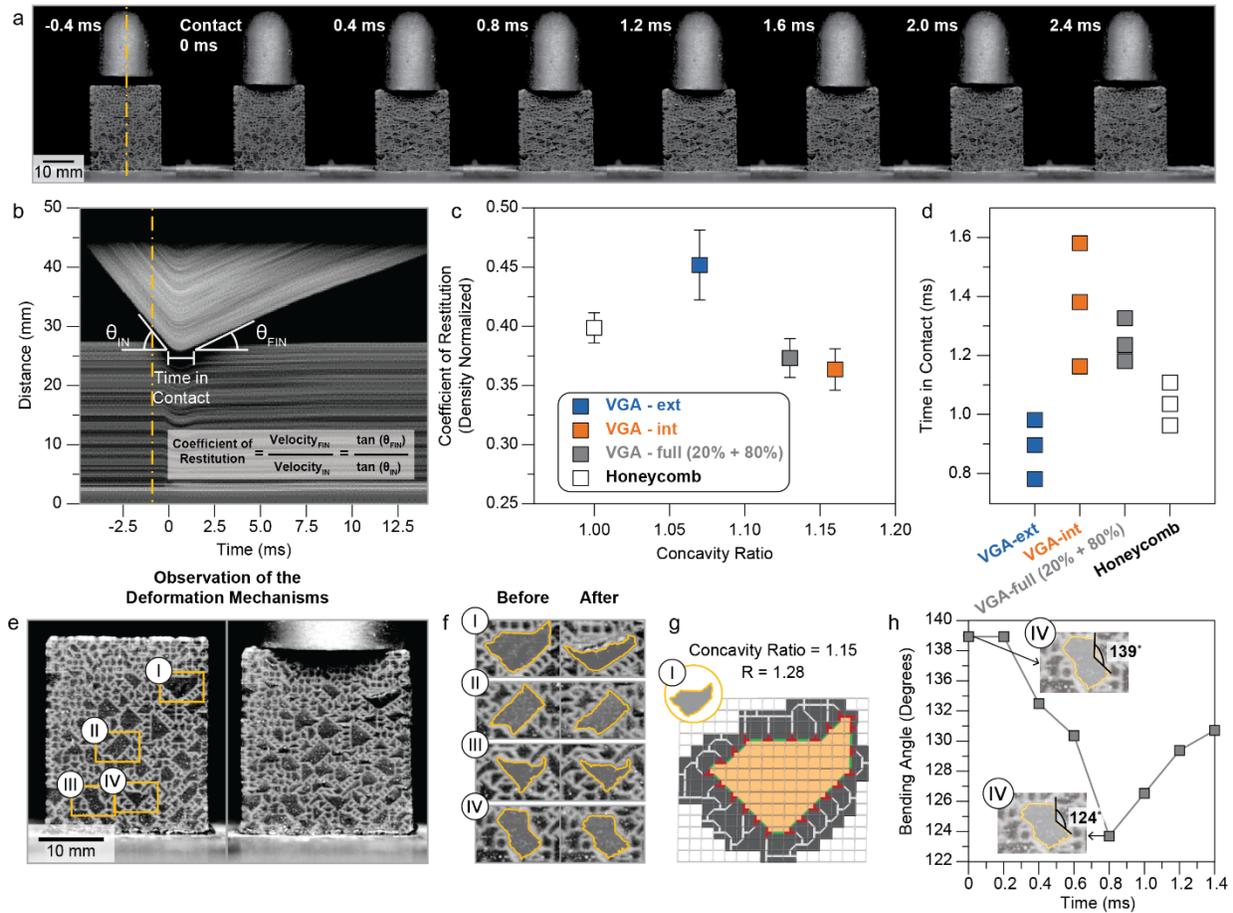

**Figure 4.** Drop tower testing. (a) VGA-full samples before, during and after loading. (b) Time versus distance of resliced striker/sample center axis with time in contact, and angles giving striker initial and final velocities. (c) Coefficient of restitution normalized for density for VGA-ext, VGA-int, VGA-full and honeycomb polymer composite samples as a function of concavity ratio. (d) Striker time in contact with sample for VGA-ext, VGA-int, VGA-full and honeycomb polymer composite samples. (e) VGA-full samples before and at maximum striker deflection with highlighted large structural features. (f) Highlighted large structural features before and after loading. (g) Coordination color coded tiling of largest structural feature. (h) Example of bending angle progression during deformation.

**Conclusions**

We present a bioinspired material design method to achieve the desirable mechanical properties of irregular biological materials in synthetic materials. The method works by reducing the complex structure of a biological material to a finite set of tiles and connectivity rules, and then using these as instructions to generate the synthetic materials. We demonstrate the method using the orange pericarp and show that the synthetic materials have similar properties to the biological material, including spatially-varying stiffness and good energy absorption, as well as isotropy as a result of irregularity. By quantifying how the tiles and connectivity rules are locally related to the mechanical performance, we also determine how to spatially control particular mechanical properties. Although this article only explores how to imitate the irregular structure of the orange pericarp and its mechanical performance, the building block method is easily extendable to imitate other irregular biological materials with desirable mechanical properties in 2D and 3D. Furthermore, this method lends itself to spatially-controlled bioinspired materials that combine the tiles and connectivity rules of multiple different materials simultaneously, to locally and globally tailor mechanical properties.

**Acknowledgements**

The authors thank Petros Arakelian, Kate Ainger, Alexander Groetsch, Lorenzo Valdevit and Adilson Motter for helpful discussions. The authors acknowledge MURI ARO W911NF-21-S-0008 for the financial support. T.M. acknowledges the Swiss National Science Foundation for the financial support.

**Conflict of Interest**

The authors declare no conflict of interest.

**Data Availability Statement**

The data that support the findings of this study are available from the corresponding authors upon reasonable request.

**Supporting Information of:**

**Extracting Geometry and Topology of Orange Pericarps for the Design of Bioinspired Energy Absorbing Materials**

*Chelsea Fox[1], Kyle Chen[1], Micaela Antonini[2], Tommaso Magrini[1,3]\* and Chiara Daraio[1]\**

[1] Division of Engineering and Applied Science, California Institute of Technology, Pasadena, CA 91125, USA

[2] Department of Biotechnology and Life Sciences, University of Insubria, Varese, Italy

[3] Department of Mechanical Engineering, Eindhoven University of Technology, 5600MB Eindhoven, The Netherlands

\* Email: t.magrini@tue.nl, daraio@caltech.edu

**Legend**

Experimental Section

Supporting Information Figures: Figures S1- S6

Supporting Information Discussion 1- 2

Supporting Information References

**Experimental Section**

*Sample Preparation and Imaging* | Tissue samples of approximately 1 cm$^3$ are collected directly from the orange fruit and include the outer layer (*exocarp* or *flavedo*) and the inner white layer (*mesocarp* or *albedo*)[1]. The samples are first immersed in a fixative solution, a 4% buffered glutaraldehyde solution at pH 7.2 (*Glutaraldehyde Solution 25% in H$_2$O*, Sigma Aldrich). The samples remain in the fixative solution for 24 hours, to ensure the complete penetration of the fixative in every part of the biological tissue. The tissues are then sequentially immersed in 70%, 80%, 90%, 95% and 100% v/v ethanol solutions (*Ethanol Absolute Anhydrous for analysis RPE*, Carlo Erba Reagents), to ensure the complete removal of any water trace. Subsequently, the tissues are immersed in terpene (natural origin, *Bio Clear*, Bio-Optica) for two hours. Terpene is an organic solvent that is miscible with alcohol and with most waxes. The next step is the infiltration of the tissues with an embedding medium. For the purpose, we transfer the tissues into disposable PVC molds, (*DispoMold*, Bio-Optica) fill the molds with liquid paraffin wax at 56°C (*Histosec Pastilles*, Sigma-Aldrich) and leave it overnight in a temperature-controlled oven at 56°C, to ensure complete infiltration of the paraffin within the tissues. The precise positioning of the tissues is critical to determine the subsequent imaging cross section. The following day, a dedicated support for histological inclusion (*Ring*, Bio-Optica) is placed on top of the PVC mold, filled with liquid paraffin at 56°C and left it to solidify at room temperature. The paraffin block is then mounted on a microtome (RMC Products, Boeckeler MR3) and thin sections are cut (~5 µm). The thin sections are then transferred on microscopy glass slides. After drying, the paraffin-embedded thin tissue slices are treated in terpene (natural origin, *Bio Clear*, Bio-Optica) to eliminate any trace of paraffin. Then we proceed with the re-hydration of the tissues. This is done by immersing the thin tissue slices in solutions with decreasing concentrations of ethanol (*Ethanol Absolute*

*Anhydrous for analysis RPE*, Carlo Erba Reagents): 100% (pure ethanol), 95%, 80%, 50%, and 0% v/v (pure water). After the re-hydration of the tissues, we proceed with the staining. First, we immerse the tissues in a 3% acetic acid solution (*Acetic Acid Solution,* Sigma Aldrich) for 5 minutes. We then perform the staining for 30 minutes using Alcian Blue (*Alcian Blue Solution,* Sigma Aldrich). After rinsing with distilled water, we then perform the counterstaining using Safranin (*Safranin T RS-for Microscopy, Hydroalcoholic Solution,* Carlo Erba Reagents) for 5 minutes. We finally wash the stained tissues using distilled water for 1 minute. Before imaging, the tissues are de-hydrated by sequential immersions in 80%, 95% and 100% v/v ethanol solutions and immersed in terpene (natural origin, *Bio Clear*, Bio-Optica) for ten minutes. Finally, the thin tissue slices are protected by a mounting cover glass, fixed on top of the microscopy glass slide with a refractive index matching acrylic resin (*Eukitt*, Bio-Optica). An upright optical microscope (Axiophot Microscope, Zeiss) is used for observation with a magnification of 10X and 20X. High resolution images were then acquired with a digital camera (CMOS 11 Discovery C30).

*Bioinspired Sample Generation* | To generate samples, we use a virtual growth algorithm developed by Liu et a.[2] and further described by Magrini et al.[3], to generate a PNG file, which is modified in Adobe Illustrator to match the volume fraction of the average orange pericarp cell wall thickness. The PNG file is then extruded and converted to an STL file to be printed using a Stratasys Objet500 Connex polyjet printer, with a stiff viscoelastic resin (Stratasys VeroWhite Polyjet Resin) for the reinforcing structure, and a soft elastomeric resin (Stratasys TangoBlack Polyjet Resin) for the matrix, both of whose mechanical properties fall within those reported in literature[4–6]. To investigate the spatially-varying density of the orange pericarp, samples with 0% external and 100% internal regions, 10% external and 90% internal regions, 20% external and 80%

internal regions, and 50% external and 50% internal regions, and 100% external and 0% internal regions were all generated. Although the actual orange pericarp has approximately the 10% and 90% region breakdown, the sample dimensions required for testing limits the size of the samples, while the printer's lateral resolution of 40-85 μm[7] limits the number of tiles that can be printed and thus the 20% and 80% sample is the optimal tradeoff to maintain a thin external region while having sufficient structural features to understand the structure-property relationship.

*Quasi-static Mechanical Testing* | Polymer composite samples with a height of 2.5 cm, a width of by 2.5 cm and a thickness of 1 cm for the external and internal regions, and with a height of 3 cm, a width of by 2.5 cm and a thickness of 1 cm for the combination samples, are tested using an Instron E3000 (Instron, USA) with a 5kN load cell and compression platens to apply compressive loading at a rate of 1 mm/min up to 10% strain.

*Digital Image Correlation* | For the digital image correlation, polymer composite samples are spray painted with flat white paint and then speckled using flat black paint to achieve speckles with a diameter of 0.1-0.3 mm. VIC 2D (Correlated Solutions, USA) is used to analyze the Lagrangian strain fields using a subset size of 35 and a step size of 2, such that each subset contains 3x3 to 5x5 speckles and each speckle contains at least 3x3 pixels.

*Dynamic Testing* | The drop tower consists of an aluminum tube with wall thickness of 0.3 cm, inner diameter of 2.7 cm, and height of 3 m, attached to a steel frame. A 10 cm steel striker with a diameter of 2.5 cm and a mass of ~400 g is released and falls down the tube to impact the sample, positioned and glued at the base, with an impact velocity of 7-8 m/s. A photodiode placed at the

base of the tube, just above the sample, captures the moment the striker passes the bottom of the tube, and is used to trigger a high speed camera (Phantom v1610, Vision Research, AMETEK, USA), which captures images of the event at 50,000 frames per second, with resolution of 512 by 512 pixels.

*Dynamic Wave Speed Testing* | The elastic wave speed is measured during the drop tower test by placing a vertical line of white dots on a sample, in the direction of loading, and using a high-speed camera (Phantom v1610, Vision Research, AMETEK, USA) at 500,000 frames per second to capture when the dots move relative to one another as a measure of when the wave arrives and then dividing the distance between dots by the time interval between arrival events (SI Figure 6).

**Supporting Information Figures**

**Figure S1**

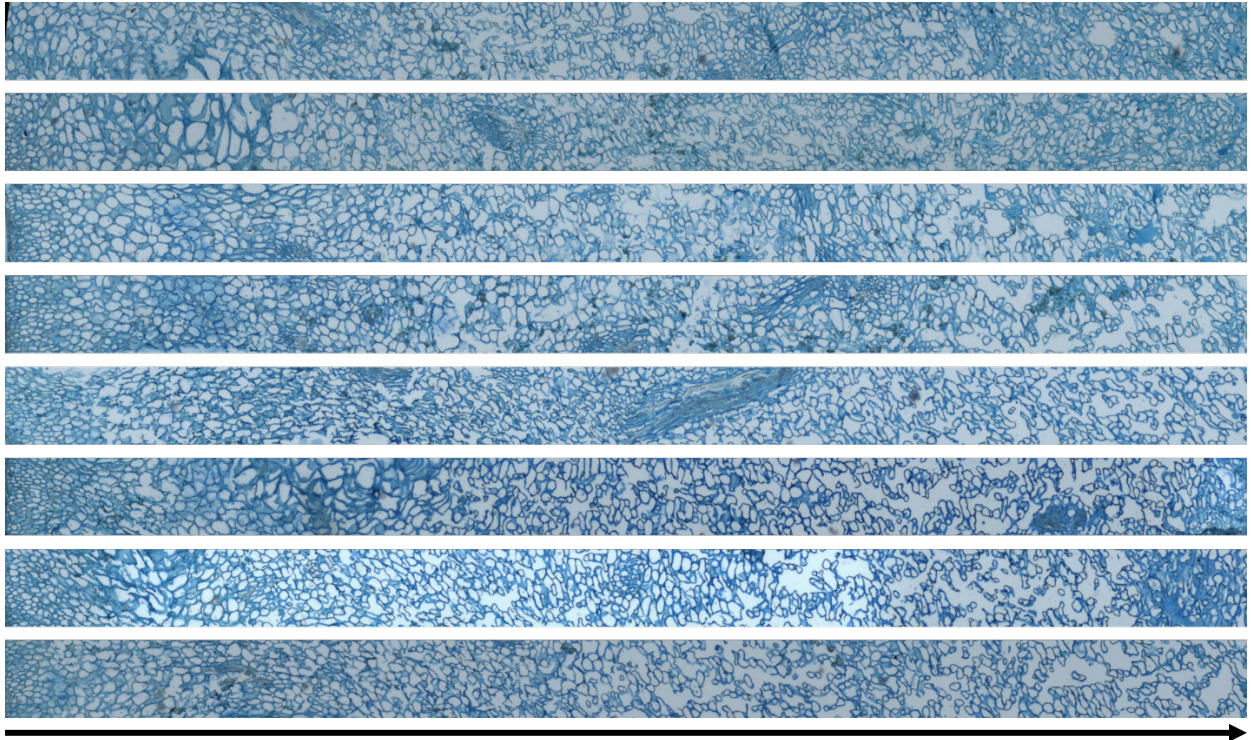

Figure S1. All orange pericarp cross sections from 0 mm to 5 mm into the fruit.

**Figure S2**

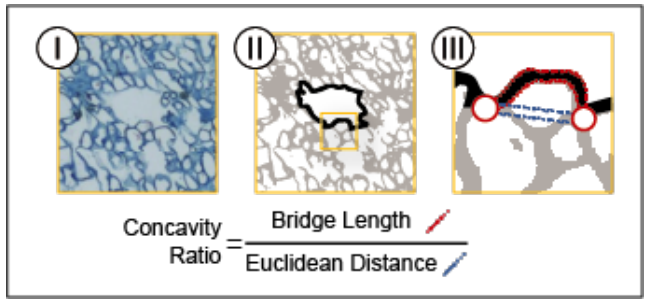

Figure S2. Concavity description. (I) Original orange pericarp image. (II) Selected feature. (III) Highlighted bridge length (red) and Euclidean distance (blue).

**Figure S3**

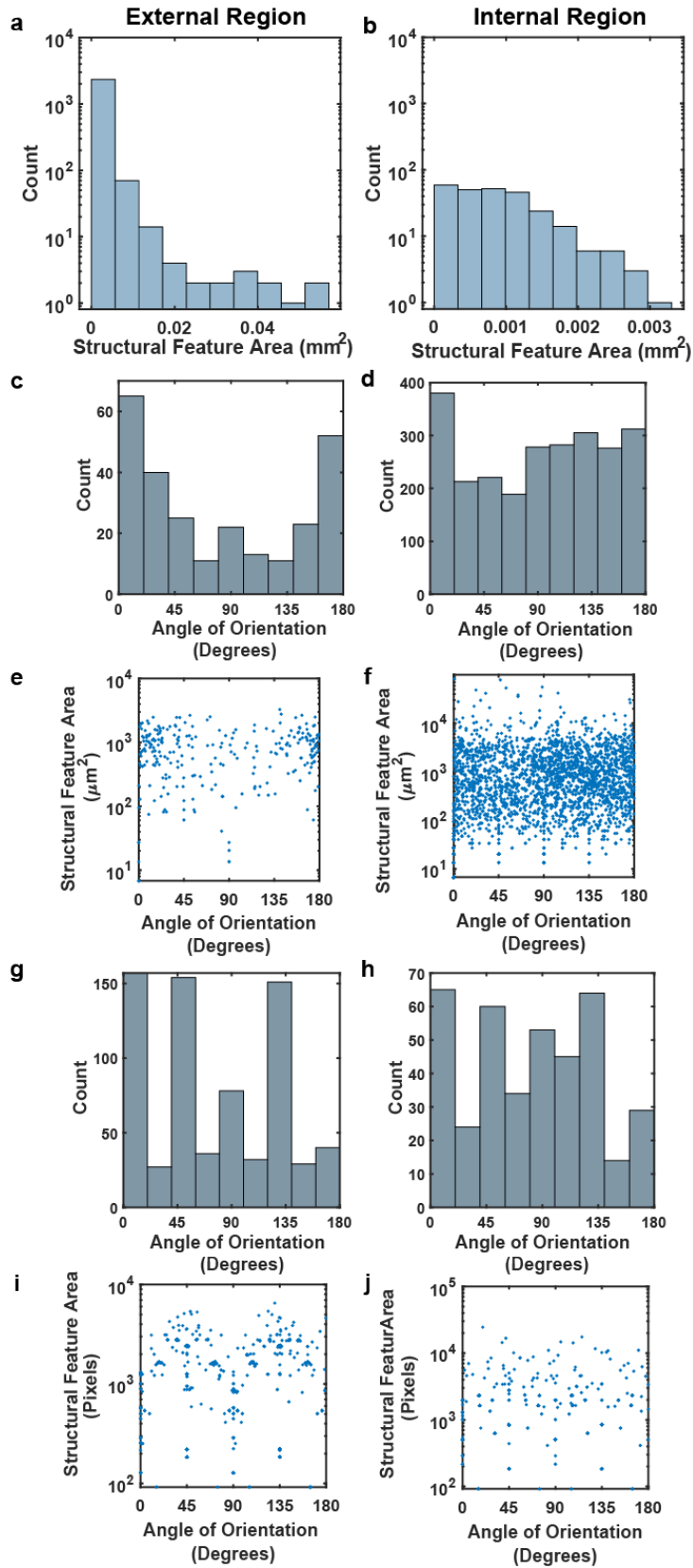

Figure S3. Orange pericarp and VGA sample characterization. (a) Counts of structural feature areas for external region of orange pericarp. (b) Counts of structural feature areas for internal region of orange pericarp. (c) Counts of structural feature angle of orientation for external region of orange pericarp. (d) Counts of structural feature angle of orientation for internal region of orange pericarp. (e) Structural feature area plotted as a function of elliptic fit angle of orientation of structural features for external region of orange pericarp. (f) Structural feature area plotted as a function of elliptic fit angle of orientation of structural features for internal region of orange pericarp. (g) Counts of structural feature angle of orientation for external region of VGA sample. (h) Counts of structural feature angle of orientation for internal region of VGA sample. (i) Structural feature area plotted as a function of elliptic fit angle of orientation of structural features for external region of VGA sample. (j) Structural feature area plotted as a function of elliptic fit angle of orientation of structural features for internal region of VGA sample.

**Figure S4**

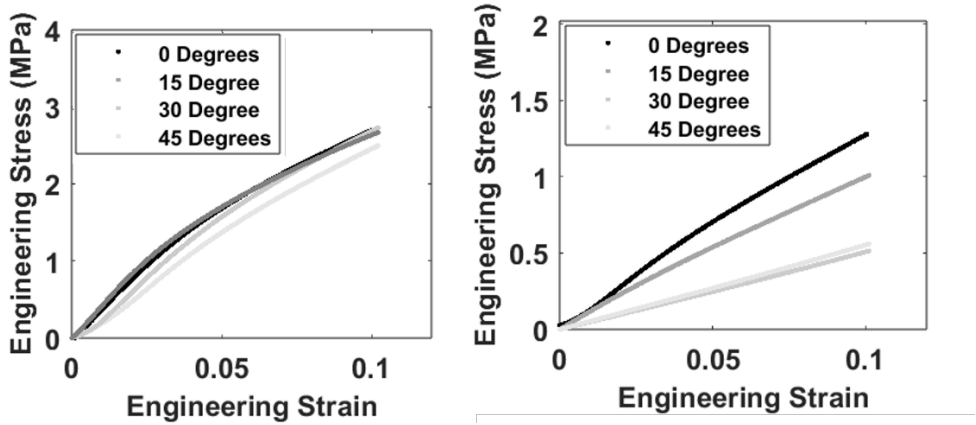

Figure S4. External region (left) and internal region (right) engineering stress-strain plot for 0-, 15-, 30-, and 45-degree rotated VGA samples.

**Figure S5**

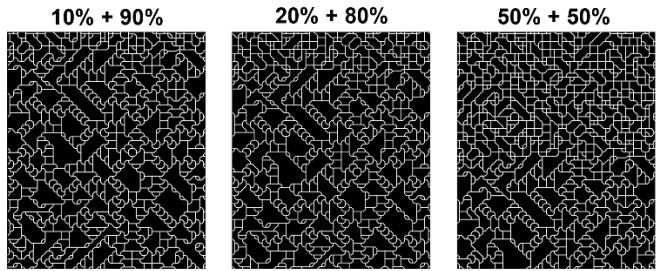

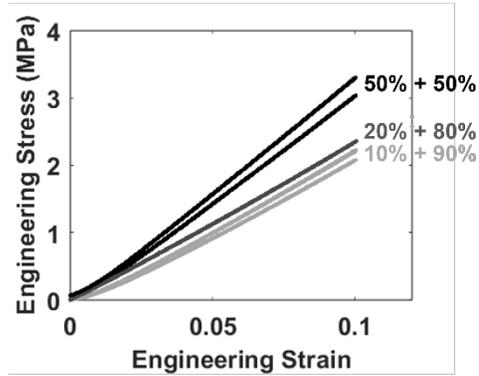

| Sample | Coefficient of Restitution |
|---|---|
| 1 - 10% + 90% | 0.29 |
| 2 - 20% + 80% | 0.33 |
| 3 - 50% + 50% | 0.36 |

Figure S5. Sample external and internal region combinations. Sample PNG's for 10% + 90%, 20% + 80%, and 50% + 50% (top). Compressive stress strain data for each variation (center). Coefficient of restitution for each variation (bottom).

**Figure S6**

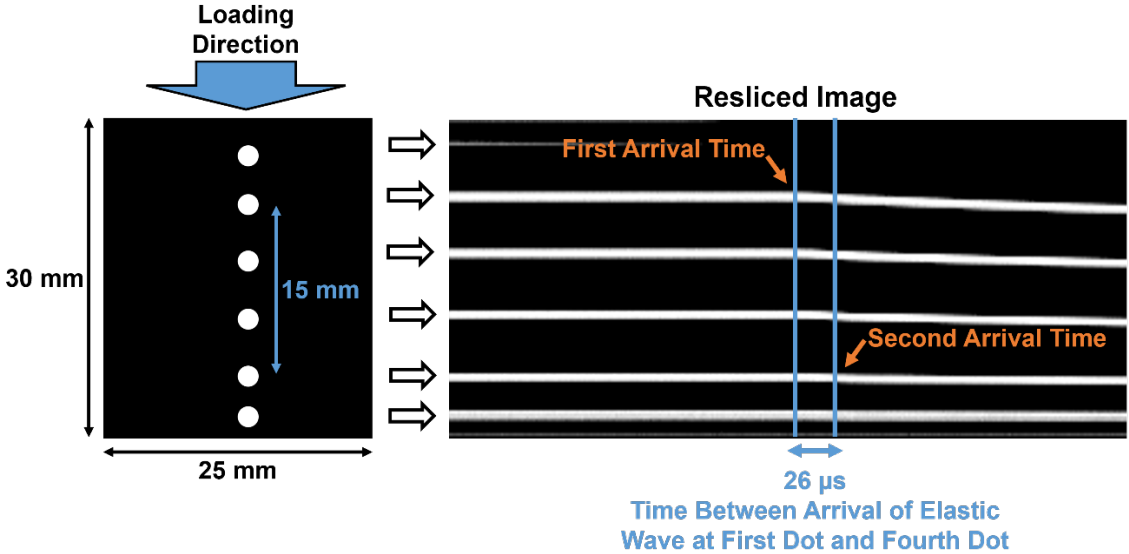

Figure S6. Wave speed propagation in composite materials. Sample with linear dot pattern and loading direction (left). Resliced image in time with dot trajectories over time as deformation reaches each dot (right).

**Supporting Information Discussions**

**Supporting Information Discussion 1 | Tiles and Connectivity Rules**

We determine the type of tile by examining its perimeter and assigning a binary value of 1 to the left, 10 to the top, 100 to the right, and 1000 to the left and summing the perimeter values for each tile. If we have a 0-coordination tile, also called 'empty tile', the perimeter values sum to 0. 2-coordination tiles display perimeter values that sum either to 101 or 1010 for tiles that feature a vertical or horizontal straight line, or to 11, 110, 1100 or 1001 for tiles that feature a diagonal line in their 4-fold rotations. 3-coordination tiles have perimeter values that sum to 111, 1011, 1110 or 1101, depending on their orientation, while 4-coordination tiles have a perimeter sum that is always equivalent to 1111. Following this counting approach, we can then assign a reduced geometry tile equivalent to the original irregular tile. It should be noted that the irregularity of the orange pericarp means that there is no preferential orientation for any tile and all orientations are approximately equally represented in the tile counts.

We determine the connectivity rules by determining how often a rule appears relative to the rest of the rules, by examining the tiles adjacent to the right and bottom of each tile. For example, if a tile with 4-coordination frequently appears next to a 3-coordination tile, such as in the external region, we allow this connectivity rule, while if a 4-coordination tile rarely appears next to 3-coordination tile, such as in the case of the internal region, we remove this connectivity rule, and these connectivity rule modifications give us the desired geometry of a region. Of course, due to the irregular nature of the original orange pericarp, all connectivity rules occur at some point in all regions, and thus we have to have a threshold for when to keep or remove a rule. If the connectivity rule appears less than 5% of the time, it is automatically removed, while if it appears less than

15%, it is removed only if either of the two tile coordination types that make up the rule appear in other connectivity rules. This ensures that the tile coordination type will still have sufficient other rules for the entire sample to be generated with the correct tile frequencies, a metric by which we verify accuracy of the generated sample to the original orange pericarp structure.

**Supporting Information Discussion 2 | Quasi-static Sample Testing**

As discussed in the main article, it should be noted that as a result of the four-sided nature of the virtual growth algorithm tiles and limited sample size, the structural features of the VGA-generated samples have a certain degree of anisotropy with respect to the amount of reinforcing material in the loading direction, as defined by feature orientation (Figure SI 3h, 3g). To investigate the effect of anisotropy on the mechanical properties of the samples, we generated VGA-ext and VGA-int samples that feature a structure rotated by 15°, 30°, and 45° and tested them in compression to determine their stiffness (Figure S4). The VGA-ext samples are nearly isotropic, with all rotated samples falling within 9% of the original sample's stiffness, while the VGA-int samples display up to 31% lower stiffness at 30° compared to 0°, indicating a higher degree of anisotropy (Figure S4). However, this is an artefact of the finite sample size that we tested and is attributed to the largest structural features with a limited number of orientations that dominate the loading response in a given sample.

**Supporting Information References**